\documentclass[prb,aps,twocolumn]{revtex4}
\usepackage{amsmath,amssymb,amstext,amscd,amsthm,amsopn,graphicx,indentfirst,mathrsfs}
\usepackage{stmaryrd}

\graphicspath{{.}{./EPS/}}

\begin{document}

\title{Alternating Superconductor--Insulator Transport Characteristics in a
Quantum Vortex Chain}
\author{Yeshayahu Atzmon$^{1}$ and Efrat Shimshoni$^{1}$ }
\affiliation{$^{1}$Department of Physics, Bar-Ilan University, Ramat-Gan 52900, Israel }

\begin{abstract}
Experimental studies of magnetoresistance in thin superconducting strips
subject to a perpendicular magnetic field $B$ exhibit a multitude of
transitions, from superconductor to insulator and vice versa alternately.
Motivated by this observation, we study a theoretical model for the
transport properties of a ladder--like superconducting device close to a
superconductor--insulator transition. In this regime, strong quantum
fluctuations dominate the dynamics of the vortex chain forming along the
device. Utilizing a mapping of the vortex system at low energies to
one-dimensional (1D) Fermions at a chemical potential dictated by $B$, we
find that a quantum phase transition of the Ising type occurs at critical
values of the vortex filling, from a superconducting phase near integer
filling to an insulator near $1/2$ --filling. The current--voltage ($I-V$)
characteristics of the weakly disordered device in the presence of a d.c.
current bias $I$ is evaluated, and investigated as a function of $B$, $I$,
the temperature $T$ and the disorder strength. In the Ohmic regime ($I/e\ll T
$), the resulting magnetoresistance $R(B)$ exhibits oscillations similar to
the experimental observation. More generally, we find that the $I-V$
characteristics of the system manifests a dramatically distinct behavior in
the superconducting and insulating regimes.
\end{abstract}

\maketitle

\section{Introduction}

\label{sec:intro}

In superconducting (SC) systems of reduced dimensionality (i.e., thin films
and wires), transport properties are strongly affected by fluctuations in
the superconducting order parameter. The most prominent manifestation of the
role of fluctuations is the appearance of a finite dissipative resistance
below the mean--field critical temperature $T_c$ of the bulk superconductor.
This failure of the hallmark of superconductivity -- the zero-resistance
character -- may persist to very low temperatures $T\ll T_c$, where pair
breaking is negligible and the electronic state can still be described in
terms of complex order parameter field representing the Bosonic degrees of
freedom. In this regime, while fluctuations in the amplitude of the order
parameter are suppressed, fluctuations in the \textit{phase} field play a
dominant role. In particular, when topological defects (vortices and
phase--slips) develop dynamics, a dissipative voltage is generated in
response to a current bias. In the $T\rightarrow 0$ limit, their quantum
dynamics dominates and may lead to the formation of a liquid phase,
characterized by a metallic or insulating behavior of the electronic system
\cite{SITrev,SIT1D}.

In the one--dimensional (1D) case, i.e. in SC wires of width and thickness
smaller than the coherence length $\xi$, the resistance essentially never
vanishes at finite $T$ due to thermal activation of phase--slips \cite%
{LAMH,TAPS} (for $T\lesssim T_c$) or their quantum tunnelling at lower $T$
\cite{Giordano,zaikin,SIT1D}. In contrast, in the two-dimensional (2D) case
(SC films), superconductivity is well-established at sufficiently low $T$.
However, by tuning an external parameter which leads to proliferation of
free vortices, it is possible to drive a quantum ($T\rightarrow 0$)
superconductor--insulator transition (SIT) \cite{QPT2,SITrev}. Employing the
concept of charge--flux duality \cite{Fisher}, one may relate the
conduction properties of the electronic system to the various phases of
vortex matter by interchanging the roles of current and voltage. Thus the SC
phase is associated with a vortex solid, while the insulator can be viewed
as a vortex superfluid.

Experimentally, one of the most convenient ways to induce a tunable SIT in
SC films is by application of a perpendicular magnetic field $B$. At fixed $T
$, a positive magnetoresistance $R(B)$ is typically observed in a wide range
of $B$. The SIT is then identified in the data as a crossing point of
these isotherms at a critical field $B_c$, separating a SC phase for $B<B_c$
from an insulating phase for $B>B_c$. At finite $T$, in both phases the
resistance is typically finite, and the distinction between the phases is
deduced from the trend of $R$ vs. $T$: $dR/dT>0$ indicates a superconductor,
and $dR/dT<0$ an insulating behavior.

Recent experimental studies of InO devices characterized by a strip geometry
\cite{shahar} -- namely, a SC wire of width comparable to $\xi$ -- offer an
opportunity to probe the crossover from a 1D to 2D quantum dynamics of the
topological phase--defects. The prominent observation is that in the
presence of a perpendicular field $B$, the magnetoresistance $R(B)$ exhibits
oscillations which amplitude is sharply increasing at low $T$, in striking
resemblance to the behavior of Josephson arrays \cite{glazman} and SC
network systems \cite{networks}. Moreover, the SIT at a high field $B_c$
appears to be preempted by a multitude of transitions at lower fields, from
a SC to an insulator or vice versa alternately. These are indicated by
multiple crossing points between different isotherms $R(B)$.

The periodicity of the above mentioned oscillations is consistent with a
single flux penetration to the sample. This suggests that the observed SC or
insulating behavior of the system is determined by commensuration of
vortices within the strip area. In particular, when an integer number of
vortices can be fitted along the strip forming a uniformly-spaced chain,
superconductivity may be supported even at sufficiently high $B$ such that a
large fraction of the sample area turns normal. However, deviation from
commensurability of the vortex filling forces a frustrated vortex
configuration, thereby weakening superconductivity. In this case, the
quantum mechanical character of vortices is manifested by the formation of
delocalized vortex states, facilitating their mobility across the width of
the strip \cite{PGR}. As a consequence, the tuning of vortex filling away from
commensurability can possibly induce a quantum phase transition to a liquid
state, of a metallic \cite{glazman} or insulating character. This
commensurate--incommensurate effect may also be manifested as magnetization
plateaux, as was predicted in a theoretical study of bosonic ladders \cite%
{OG1}.

\begin{figure}
\includegraphics[width=1.0\linewidth]{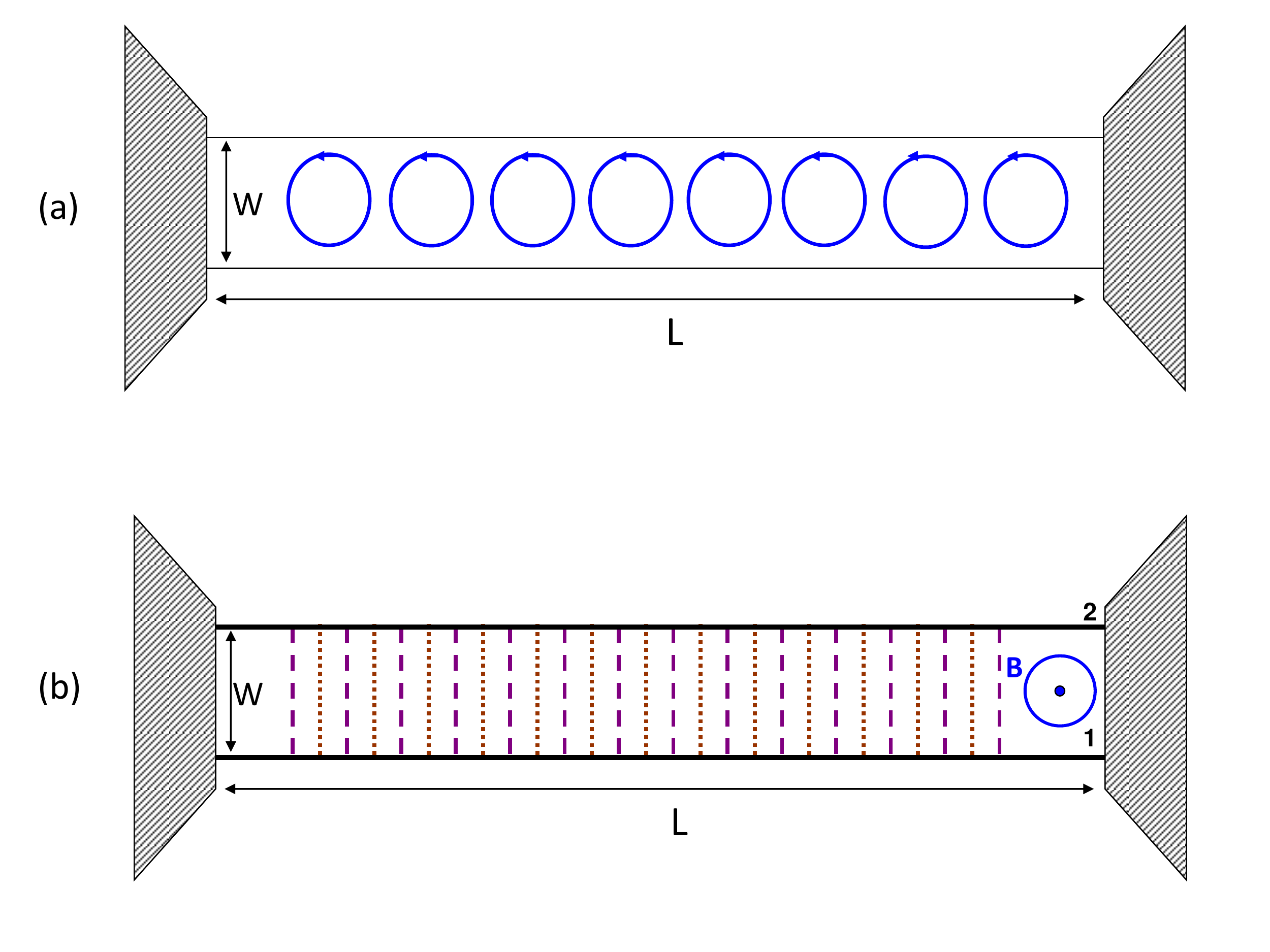}
\caption{(color online)
(a) Top view on a chain of vortices in a superconducting strip. (b) The line-junction model for the system; purple dashed lines represent the Josephson coupling $g_J$, and brown dotted lines the Coulomb interaction $U$ between the two SC wires.
  \label{device} }
\end{figure}

In a recent paper \cite{AS} we have studied this phenomenon within a
theoretical model for a quantum vortex chain in a ladder-like SC device (see Fig. \ref{device}),
which particularly addresses the strongly quantum fluctuation regime where
the parameters are close to a SIT. It was shown that such system may exhibit
multiple quantum phase transitions of the Ising type, manifested as
SC--insulator oscillations of the Ohmic resistance $R(T,B)$. This reflects
an intimate correspondence between charge-flux duality across a SIT, and the
order-disorder duality characterizing the Ising transition at $1+1$%
-dimensions.

In this paper we present a detailed theory for the electric transport
properties of the quantum vortex chain in a weakly disordered SC ladder. In
particular, we derive the current--voltage ($I-V$) characteristics of the
device in the presence of a d.c. current bias $I$, and investigate their
behavior as a function of $B$, $I$, the temperature $T$ and the disorder
strength. We find that the $I-V$ characteristics of the system manifests a
dramatically distinct behavior in the SC and insulating regimes. In the
Ohmic regime ($I/e\ll T$), this yields an oscillatory magnetoresistance $%
R(T,B)$ which exhibits $T$--dependence compatible with the experimental data.

The paper is organized as follows: in Sec. \ref{sec:model} we construct the
line--junction model for the SC strip, and derive its mapping to 1D Fermions
and consequently to the quantum Ising chain. In Sec. \ref{sec:transport} we
provide a detailed calculation of the dissipative voltage in a
current--biased strip, and derive expressions for the non-linear $I-V$
characteristics and $T$--dependent magnetoresistance in the various regimes
(the SC phases, insulating phases and critical regions). Our conclusions and
discussion of the relation to further experiments are summarized in Sec. \ref%
{sec:summary}.

\section{The Model}

\label{sec:model}

We consider a thin SC strip of length $L\gg \xi $ and width $w\gtrsim \xi$,
subject to a strong perpendicular magnetic field below the 2D SIT (i.e., $%
B\lesssim B_c$). A 1D chain of vortices is formed along the central axis of
the strip, which can be viewed as a 1D system of particles in the presence
of a self--organized effective potential dictated by the combination of
vortex-vortex interaction and the boundary conditions [Fig. \ref{device}(a)]. In particular, the
interface with the vacuum at the strip edges induces an effective "image
charges" potential \cite{likharev}, and bulk-superconductor contacts
connected to both ends of the strip enforce a fixed phase of the SC order
parameter at $x=\pm L/2$. As a result, the effective potential acquires the
form of a periodic 1D lattice of pinning sites separated by a uniform
spacing $a=L/N$, where $N={\mathcal{I}}[BwL/\Phi _{0}]$ (with $\Phi
_{0}=hc/2e$ the flux quantum, and ${\mathcal{I}}[z]$ the integer value of $z$%
) denotes the total number of vortices \cite{glazman}. Assuming further that
the high vortex density in this case leads to near merging of their cores
along the central axis of the strip, the system becomes essentially
equivalent to a line--junction formed by a pair of parallel SC wires
separated by a normal barrier [Fig. \ref{device}(b)], subject to a magnetic field $B$ perpendicular
to the junction plane.

In the low $T$ regime, pair-breaking is negligible and the properties of
this system are dominated by quantum phase-fluctuations of the SC
condensate. It is therefore possible to model it as a 2--leg bosonic ladder
\cite{OG1} (or, equivalently, a ladder-like Josephson array \cite{glazman}),
where a coordinate $x=ja$ ($j$ integer) denotes the locations of vortex
cores in the continuum limit. The dynamics of the collective phase field in
the wires ($\phi _{n}(x,t)$ with $n=1,2$) is governed by the effective 1D
Hamiltonian
\begin{equation}
H_{0}=H_{1}+H_{2}+H_{int}\;,  \label{Hfull}
\end{equation}
in which (using units where $\hbar =1$)
\begin{eqnarray}
H_{n} &=&\frac{1}{2}\int_{-\frac{L}{2}}^{\frac{L}{2}}dx\left[ U_{0}\rho
_{n}^{2}+\frac{\rho _{s}}{4m}(\partial _{x}\phi _{n})^{2}\right] \;,
\label{H_i} \\
H_{int} &=&\int_{-\frac{L}{2}}^{\frac{L}{2}}dx\left[ -g_{J}\cos (\phi
_{1}-\phi _{2}-qx)+U\rho _{1}\rho _{2}\right] .  \label{Hint}
\end{eqnarray}
Here the operator $\rho _{n}(x)$ denotes density fluctuations of Cooper
pairs in wire $n$, and can be represented as \cite{book}
\begin{equation}
\rho _{n}(x)=-\frac{1}{\pi }\partial _{x}\theta _{n}(x)+\rho
_{0}\sum_{p\not=0}e^{i2p(\pi \rho _{0}x-\theta _{n})}  \label{rhototheta}
\end{equation}
in terms of the conjugate field $\theta _{n}(x)$ satisfying $[\phi
_{n}(x),\partial _{x}\theta _{n}(x^{\prime })]=i\pi \delta (x^{\prime }-x)$.
The first term in Eq. (\ref{H_i}) hence describes a charging energy; $\rho
_{s}$ is the superfluid density (per unit length) assumed to be
monotonically suppressed by increasing $B$, $\rho _{0}=\rho _{s}(B=0)$ and $%
m $ is the electron mass. The inter--wire coupling [Eq. (\ref{Hint})]
consists of a Josephson term and an inter--wire Coulomb interaction, of
coupling strengths $g_{J}$ and $U$, respectively. Finally, the parameter
\begin{equation}
q=2\pi\frac{w(B-B_N)}{\Phi_0}\; ,\quad B_N=NB_0\,,\quad B_0\equiv \frac{%
\Phi_0}{wL}  \label{new_q}
\end{equation}
parametrizes the deviation of the vortex density from the closest
commensurate value, i.e., it denotes vortex ``doping". We note that $H_{0}$
describes an ideal system, to which we later add a disorder potential.

To further analyze the properties of this model, it is convenient to
introduce symmetric and antisymmetric phase and charge fields via the
canonical transformation
\begin{equation}
\phi_\pm=\frac{1}{\sqrt{2}}(\phi_1\pm\phi_2),\,\,\,\, \theta_\pm=\frac{1}{
\sqrt{2}}(\theta_1\pm\theta_2)\; .  \label{symmetric_antisymmetric}
\end{equation}
In terms of these variables, the Hamiltonian (\ref{Hfull}) is separable:
%\begin{equation}
%H_0=H_++H_-  \label{Hpm}
%\end{equation}
\begin{widetext}
%\begin{equation}
\begin{multline}
H_0=H_++H_- \\
\mathrm{where}\quad\quad\quad  H_+ = H_{LL}^{(+)}\; ,\quad\quad\quad
\quad H_- = H_{LL}^{(-)}+
\int_{-\frac{L}{2}}^{\frac{L}{2}} dx\left[-g_J\cos(\sqrt{2}\phi_--qx) +
g_c\cos(\sqrt{8}\theta_-)\right] \; ;  \label{Hpm}
\end{multline}
%\end{equation}
\end{widetext}
\begin{equation}
 H_{LL}^{(\pm)} \equiv  \frac{v_\pm}{2\pi}\int_{-\frac{L}{2}}^{\frac{L}{2}} dx %
\left[ K_\pm (\partial_x\theta_\pm)^+\frac{1}{K_\pm} (\partial_x \phi_\pm)^2 %
\right]  \label{HLL}
\end{equation}
and the parameters are given by
\begin{eqnarray}
\quad K_\pm &=&\sqrt{\frac{4m(U_0\pm U)}{\pi^2\rho_s}}\; ,\quad v_\pm=\sqrt{%
\frac{\rho_s(U_0\pm U)}{4 m}}\; ,  \notag \\
g_c&=&2U\rho_0^2\; .  \label{Kvg_c}
\end{eqnarray}
Here we have accounted for the most relevant interaction terms, neglecting
umklapp terms included in the charging energy which are effectively
suppressed due to the rapidly oscillating factor in Eq. (\ref{rhototheta}).
The symmetric mode (corresponding to the plasmons of total charge) governed
by $H_+$ is therefore gapless. However, the behavior of the antisymmetric
mode is dictated by the competition between two interacting (cosine) terms,
and depends crucially on the value of the Luttinger parameter $K_-$. Below
we focus on the regime of parameters close to a SIT in 1D wires, where
quantum fluctuations in the phase and charge fields are maximized; i.e., $%
K_-\approx K_c=2$ (see Ref. \onlinecite{zaikin}).

We next define new canonical fields
\begin{equation}
\phi \equiv \frac{1}{\sqrt 2 }\phi _ -\; ,\quad \theta \equiv \sqrt{2}
\theta _ -  \label{new_theta}
\end{equation}
in terms of which $H_{LL}^{(-)}$ acquires the form of a Luttinger
Hamiltonian with an effective Luttinger parameter $K = K_-/2$. For $K_-$
close to $K_c=2$, we thus obtain $K\approx 1$. This yields
\begin{eqnarray}
H_- &=& \frac{v_-}{2\pi}\int_{-\frac{L}{2}}^{\frac{L}{2}} \left[ (\partial
_x \theta )^2 + (\partial _x \phi )^2 \right] \\
&+& \int_{-\frac{L}{2}}^{\frac{L}{2}}dx \left[-g_J\cos (2 \phi -qx ) +
g_c\cos (2 \theta ) \right]\; .  \notag  \label{H_-bos}
\end{eqnarray}
This model can be refermionized by introducing right ($R$) and left ($L$)
moving spinless Fermion fields \cite{SNT}
\begin{equation}
\psi _{R,L} = \frac{1}{{\sqrt {2\pi \alpha } }}e^{\pm ik_Fx}e^{i( \mp \phi +
\theta )}\; ,  \label{fermions_def}
\end{equation}
in terms of which $H_-$ becomes a free Hamiltonian. Here the short-distance
cutoff $\alpha$ is set by the lattice constant $a$ characterizing the vortex
chain, and the ``Fermi momentum" $k_F=\pi/a+q$ is determined by the vortex
filling factor [see Eq. (\ref{new_q})]. Quite interestingly, this implies
that near a SIT, it is natural to adapt a duel representation of this system
in terms of \textit{fermionic} vortex fields. This stems from the
approximate self-duality of $H_-$ (i.e., its symmetry to exchange of $\phi$
and $\theta$), implying that the natural degrees of freedom are composites
of a pair charge ($2e$) and a unit of flux quantum.

The fermionic representation of $H_-$ is given by
\begin{multline}
H_- = \int dx \{ v_ - [\psi _R^\dag (x)( - i\partial _x )\psi _R (x) - \psi
_L^\dag (x)( - i\partial _x )\psi _L (x)] \\
- \mu_v [\psi _R^\dag (x)\psi _R (x) + \psi _L^\dag (x)\psi _L (x)] \\
-J [\psi _R^\dag (x)\psi _L (x) + \psi _L^\dag (x)\psi _R (x)] \\
+ V [\psi _R^\dag (x)\psi _L^\dag (x) + \psi _L (x)\psi _R (x)]\}
\label{H_-_ferm}
\end{multline}
where $J=\pi\alpha g_J$, $V=\pi\alpha g_c$ and the vortex chemical potential
is $\mu_v=\pi v_-q$, which vanishes at commensurate fillings. Following the
analogous problem of spin-$1/2$ ladders \cite{SNT,Tsvelik}, it is useful to
decompose the complex Fermions [Eq. (\ref{fermions_def})] in terms of the
Majorana fields
\begin{equation}
\eta _{1\nu} = \frac{1}{\sqrt 2}\left(\psi_\nu+ \psi _\nu^\dag \right)\;
,\quad \eta _{2\nu} = \frac{1}{i\sqrt 2}\left(\psi_\nu- \psi _\nu^\dag
\right)  \label{majo_def}
\end{equation}
($\nu=R,L$). Recasting Eq. (\ref{H_-_ferm}) in $k$-space and using the
Fourier transformed fields $\eta_{j\nu,k}=\eta_{j\nu,-k}^\dagger$, we obtain
\begin{multline}  \label{H_-_majo}
H_ - = \sum_k \Psi_k^\dagger \mathcal{H}_k \Psi_k\; , \\
\mathcal{H}_k \equiv \left( {\begin{array}{*{20}c} {v_- k } & {
i\Delta_u^{(0)}} & {i\mu_v } & 0 \\ {-i\Delta_u^{(0)}} & { - v_- k } & 0 & {
- i\mu_v } \\ { - i\mu_v } & 0 & {v_- k } & {-i\Delta_d^{(0)}} \\ 0 &
{i\mu_v } & { i\Delta_d^{(0)}} & { -v_- k } \\ \end{array}} \right) \\
\Psi_k^\dagger \equiv \left( {\begin{array}{*{20}c} {\eta_{1R,k} } & , &
{\eta_{2L,k} } & , & {\eta_{2R,k} } & , & {\eta_{1L,k} } \end{array}}
\right)\; ;
\end{multline}
here
\begin{equation}
\Delta_{u,d}^{(0)}=J\pm V  \label{delta_ud0}
\end{equation}
denote the gaps in the excitation spectrum for commensurate vortex filling ($%
\mu_v=0$), in which case $\mathcal{H}_k$ decouples into two independent
blocks. Since $J,V$ are positive, the $u$ sector is higher in energy.

We now focus on the case of interest, where the system is assumed to be in
the SC phase but close to a SIT so that the Josephson energy $J$ is slightly
larger than $V$, and $\Delta_{d}^{(0)}\ll \Delta_u^{(0)}$. In this case, the
high energy sector $u$ can be truncated, and the low-energy properties are
governed by the $d$-type Fermions. Most notably, the gap $\Delta_{d}^{(0)}$
can \textit{change sign} upon tuning of $J$ below the critical value $J_c=V$
where $\Delta_{d}^{(0)}=0$. Indeed, for $\mu_v=0$ each species of free
massive Fermion models described by (\ref{H_-_majo}) can be independently
mapped to an Ising chain in a transverse field \cite{Ising,QPT2}. In
particular, the low energy sector $d$ can be described by the spin
Hamiltonian
\begin{equation}
H_d=-j\sum_j\sigma^z_j\sigma^z_{j+1}-V\sum_j\sigma^x_j  \label{H_Ising_d}
\end{equation}
which possesses a quantum critical point at $J=V$.

When finite vortex ``doping" is introduced by tuning $B$ away from $B_N$
such that $\mu_v\not=0$, the original $d$ and $u$ sectors mix. However, the
resulting long wave-length theory can still be cast in terms of two
decoupled sectors denoted $d$ (low) and $u$ (high). Moreover, the energy
spectrum
\begin{eqnarray}
\epsilon_{u,d}(k)&=&\left[J^2+ \tilde V^2+v_-^2k^2\pm 2\sqrt{J^2\tilde
V^2+(\mu_vv_-)^2k^2}\right]^{1/2}\; ,  \notag \\
\tilde V&\equiv &\sqrt{V^2+\mu_v^2}  \label{spectrum}
\end{eqnarray}
reduces in the $k\rightarrow 0$ limit to the same form as the $\mu_v=0$
case:
\begin{equation}
\epsilon_{u,d}(k)\approx \Delta_{u,d}+\frac{1}{2}\frac{v^2_{u,d}k^2}{%
\Delta_{u,d}}\; ,
\end{equation}
with the modified velocities
\begin{equation}
v^2_{u,d}=v_-\left(1\pm \frac{\mu_v}{J\tilde V}\right)
\end{equation}
and modified gaps given by
\begin{equation}
\Delta_{u,d}(B)=J\pm \tilde V\; .  \label{gaps}
\end{equation}
The $B$-dependence of $\Delta_{u,d}$ is oscillatory due to the dependence of
$\tilde V$ on the vortex doping $\mu_v$ [Eq. (\ref{spectrum})]. While $%
\Delta_u$ remains positive and large for arbitrary $\mu_v$, a quantum phase
transition occurs at a critical value of $\mu_v$ [which can be traced back
to a \textit{sequence} of critical fields $B_c^{(N)}$ via $\mu_v(q)$ and Eq.
(\ref{new_q})], where $\Delta_{d}$ changes sign. As $B\rightarrow B_c^{(N)}$%
, one expects the scaling
\begin{equation}
|\Delta_d|\sim |B-B_c^{(N)}|\; .  \label{gap_B}
\end{equation}
As we show in the next Section, the above discussed Ising like quantum
critical points correspond to SC--insulator transitions, marked by a
dramatic change in the transport properties.

\section{I-V Charactaristics and Magnetoresistance}

\label{sec:transport}

We next study the transport properties of the system in the presence of a
weak scattering potential, generically induced by random, uncorrelated
impurities along the coupled wires. To this end, we include a linear
coupling of the density operator $\rho _{n}(x)$ [Eq. (\ref{rhototheta})] to
a disorder potential $V_{D}(x)$ in the Hamiltonian. The leading contribution
to dissipation arises from the backscattering term of the form \cite{book}
\begin{equation}
H_{D}=\sum_{n=1,2}\int dx\zeta_n (x)\cos \{2\theta _{n}(x)\}  \label{16}
\end{equation}
where we assume
\begin{equation}
\langle \zeta_n (x)\rangle =0\; ,\quad \langle \zeta_n (x)\zeta_{n^{\prime}}
(x^{\prime })\rangle =D\delta (x-x^{\prime })\delta_{n,n^{\prime}}.
\label{def_D}
\end{equation}
Here and throughout the rest of the section, the definition of $\langle
\,\rangle $ includes disorder averaging. As a result of phase-slips
generated by $H_D$, a finite voltage will develop along the SC strip when
driven by a current bias $I$.

To introduce a d.c. current bias $I$, we add a time-dependent term $It$ to
the total charge operator
\begin{equation}
Q=-\frac{2e}{\pi}(\theta_{1}+\theta_{2}).\nonumber
\end{equation}
Using Eq. (\ref{symmetric_antisymmetric}), this yields
\begin{equation}
\theta_{+}(x,t)=-\frac{\pi}{2\sqrt{2}e}Q(x,t)=\tilde\theta_+(x,t) -\frac{\pi
}{2\sqrt{2}e} It  \label{17}
\end{equation}
where $\tilde{\theta}_+(x,t)$ describes equilibrium fluctuations ($I=0$).
The induced voltage along the strip is then given by $V\equiv\langle\hat{V}
(L/2,t)\rangle$, where the voltage operator $\hat V(x,t)$ is dictated by the
Josephson relation
\begin{eqnarray}
\hat{V}&=&\frac{1}{2e}(\dot{\phi_{1}}+\dot{\phi_{2}})= \frac{1}{\sqrt{2}e}
\dot{\phi}_{+}\; ,  \label{18} \\
\dot{\phi}_{+} &=& i[H,\phi_{+}].  \nonumber
\end{eqnarray}
Using $H=H_{0}+H_{D}$ [Eqs. (\ref{Hpm}),(\ref{16})] we find
\begin{equation}
\begin{array}{lllllllllllllllllll}
{{{\dot \phi }_ + }(x,t) = \frac{{{v_ + }{K_ + }}}{{\sqrt 2 e}}\left\{ {{
\partial _x}{\theta _ + }(x,t)} \right\}} &  &  &  &  &  &  &  &  &  &  &  &
&  &  &  &  &  &  \\
{\ - \frac{\pi }{e}\sum\limits_{n = 1,2} {\int\limits_{ - \frac{L}{2}}^x {d{
x^\prime }{\zeta _n}({x^\prime })\sin \left[ {2{\theta _n}({x^\prime },t)}
\right]\,\,} } .} &  &  &  &  &  &  &  &  &  &  &  &  &  &  &  &  &  &
\end{array}
\end{equation}

The time-evolution of $\dot\phi_+(x,t)$ can be expressed as
\begin{equation}
\dot\phi_+(t)=u(t)\dot{\widetilde{\phi}}_+(t)u^{\dagger }(t),  \label{21}
\end{equation}
where $\dot{\widetilde{\phi}}_+(t)$ is the operator in the interaction
representation
\begin{equation}
\dot{\widetilde{\phi}}_+(t)=e^{iH_0t}\dot\phi_+e^{-iH_0t}\; ,
\end{equation}
and
\begin{equation}
u(t)\equiv e^{i(H_{0}+H_{D})t}e^{-iH_{0}t}.  \label{4}
\end{equation}
Assuming a weak disorder which allows a perturbative treatment of $H_{D}$, $
u(t)$ is given to first order by
\begin{equation}
u(t)=1+{\underset{-\infty }{i\int}^t}dt^{\prime }H_{D}(t^{\prime }).
\label{25}
\end{equation}
Substituting Eq. (\ref{25}) in Eq. (\ref{21}), one obtains
\begin{equation}
\langle\dot{\phi_+ }(x,t)\rangle=i{\overset{t}{\underset{-\infty }{\int }}
dt^{\prime }\left\langle \left[ H_{D}(t^{\prime }),\dot{\widetilde{\phi}}
_+(x,t)\right] \right\rangle } .  \label{26}
\end{equation}

Using Eqs. (\ref{16}), (\ref{18}) and (\ref{26}), and recalling
Eq. (\ref{symmetric_antisymmetric}), we obtain an expression for the d.c.
voltage
\begin{equation}
V=V_{1}+V_{2}
\end{equation}
where
\begin{widetext}
\begin{equation}
V_{1 (2)}\equiv\frac{iDL\pi}{e}\int^t_{-\infty}dt^{\prime}\left\langle\left[ \sin \left(
\sqrt{2}\{\theta _+(t)\pm\theta_{-}(t)\}\right) \cos \left(\sqrt{2}\{\theta _+(t^{\prime})\pm\theta_-(t^{\prime})\}\right)\right]\right\rangle
\end{equation}
\end{widetext}
(here $\theta _{\pm }(t)\equiv \theta _{\pm }(0,t)$). Introducing the
operators
\begin{equation}
A_{1 (2)}(x,t) \equiv e^{i\sqrt{2}(\tilde{\theta}_+(x,t)\pm \theta
_{-}(x,t))}  \label{29}
\end{equation}
where $\tilde{\theta}_+$ is defined in Eq. (\ref{17}), we obtain the
voltage-current characteristic
\begin{widetext}
\begin{equation}
%\begin{array}{l}
V(I) = \frac{{DL\pi }}{{4e}}\sum_{n = 1,2}
\int_{- \infty }^\infty  d{t^\prime } i\Theta (t - {t^\prime }) \left\{
{e^{i\frac{{\pi I}}{{2e}}({t^\prime } - t)}}\left\langle {\left[ {{A_n}(t),A_n^\dag ({t^\prime })} \right]} \right\rangle
 - {e^{ - i\frac{{\pi I}}{{2e}}({t^\prime } - t)}}\left\langle {\left[ {A_n^\dag (t),{A_n}({t^\prime })} \right]}
 \right\rangle \right\}\; .
%\end{array}
   \label{30}
\end{equation}
\end{widetext}
In terms of the retarded Green's functions
\begin{equation}
\begin{array}{ll}
\chi^{(n)}_{ret}(t)&= -i\Theta (t)\left\langle \left[ A_{n}(t),A^{
\dagger}_{n}(0)\right] \right\rangle \\
&= -2\Theta(t)\Im m\{\chi_n(t)\}
\end{array}
\label{31}
\end{equation}
with
\begin{equation}
\chi_n(t)\equiv\langle A_n(t)A^\dagger_{n}(0)\rangle,  \label{32}
\end{equation}
we finally obtain
\begin{equation}
\begin{array}{ll}
V(I) & =\frac{DL\pi}{2e}\sum\limits_{n=1,2}\overset{\infty }{\underset{0}{
\int }}dt\sin \left(\frac{\pi It}{2e}\right)\Im m\{{\chi_n(t)}\} \\
&= \frac{DL\pi}{4e}\sum\limits_{n=1,2}\overset{\infty}{\underset{-\infty}{\int
}}dt\sin (\frac{\pi It}{2e})\chi_n(t)
\end{array}
\label{33}
\end{equation}
%\end{widetext}
where in the last step we have used the fact that $\Im m\{\chi_n(t)\}$ is the
antisymmetric part of $\chi_n(t)$ under $t\shortrightarrow -t$. This correlation function can be evaluated utilizing the low-energy theory developed in Sec. \ref{sec:model}.

To leading order in the perturbation $H_D$, the expectation value $\langle\,\rangle$ may be
replaced by $\langle\,\rangle_0$, evaluated with respect to $H_0$. Since
the ${\theta_+}$, $\theta_-$ degrees of freedom are decoupled in $
H_0$,
%any correlation function of the form $\langle f(\tilde{\theta}
%_+)g(\theta_-)\rangle_0=\langle f(\tilde{\theta}_+)\rangle_0\langle
%g(\theta_-)\rangle_0$. Hence,
the correlation function
\begin{equation}
\chi_1=\chi_2\equiv\chi
\end{equation}
where
\begin{equation}
\chi (t)=\left\langle e^{i\sqrt{2}(\tilde{\theta}_+(x,t)+\theta
_{-}(x,t))}e^{-i\sqrt{2}(\tilde{\theta}_+(0,0)+\theta
_{-}(0,0))}\right\rangle_0  \label{34}
\end{equation}
\ \ \ \ decouples into
\begin{widetext}\begin{eqnarray}
\chi (t) &=&\chi _{C+}(t)\chi _{C-}(t)+\chi_{S+}(t)\chi_{S-}(t)
+\chi_{S+}(t)\chi _{C-}(t)+\chi _{C+}(t)\chi _{S-}(t)  \\
\chi _{C\pm } &\equiv &\langle \cos \{\sqrt{2}\theta _{\pm }(t)\}\cos \{
\sqrt{2}\theta _{\pm }(0)\}\rangle _{\pm }\; ,\quad\quad
\chi _{S\pm } \equiv \langle \sin \{\sqrt{2}\theta _{\pm }(t)\}\sin \{
\sqrt{2}\theta _{\pm }(0)\}\rangle _{\pm }\; .
\end{eqnarray}\end{widetext}
Here $\langle \,\rangle _{\pm }$ are evaluated with respect to $H_{\pm}$. The symmetric mode described by $H_{+}$ is a Luttinger liquid [see Eq. (\ref{Hpm})], hence
\cite{book}
\begin{equation}
\chi_{C+}(t)=\chi_{S+}(t)=\lim_{\epsilon\rightarrow 0}\left(\frac{
-(\pi\alpha T/v_+)}{\sinh\{\pi T\left(t-i\epsilon\right)\}}\right)^{\frac{1}{
K_+}}.  \label{chi_+}
\end{equation}
In contrast, as discussed below, the correlations characterizing the antisymmetric mode [$\chi_{C-}(t)$ and $\chi_{S-}(t)$] depend
crucially on the parameters of (\ref{H_-_majo}), and in particular on the
magnitude and \textit{sign} of the masses $\Delta_{u,d}$.

To evaluate $\chi_{C-}$ and $\chi_{S-}$, we first note that in terms of the
field $\theta$ [Eq. (\ref{new_theta})], they correspond to correlation
functions of $\cos\theta$, $\sin\theta$, which lack a local representation
in terms of Fermion fields. However, a convenient expression is available in
terms of the two species of order ($\sigma_{u,d}$) and disorder
($\tilde\sigma_{u,d}$) Ising fields \cite{SNT,Ising}: for $\Delta_d>0$,
\begin{equation}
\cos\theta\sim \sigma_u\tilde\sigma_d\; ,\quad \sin\theta\sim
\tilde\sigma_u\sigma_d\; .  \label{Ising_udp}
\end{equation}
For $\Delta_d<0$, the roles of $\sigma_d$, $\tilde\sigma_d$ are simply
\textit{interchanged}.
%whereas for $\Delta_d<0$, the roles of $\sigma_d$, $\tilde\sigma_d$ are interchanged:
%\begin{equation}
%\cos\theta\sim \sigma_u\sigma_d\; ,\quad \sin\theta\sim \tilde\sigma_u\tilde\sigma_d\; .
%\label{Ising_udm}
%\end{equation}
The correlators $\chi_{C-}$, $\chi_{S-}$ can therefore be expressed in terms
of $C_\lambda(t)=\langle \sigma_\lambda(t)\sigma_\lambda(0)\rangle$, $\tilde
C_\lambda(t)=\langle \tilde\sigma_\lambda(t)\tilde\sigma_\lambda(0)\rangle\,$
($\lambda=u,d$), which have known analytic approximations in the
semi--classical regime ($|\Delta_\lambda|\gg T$) \cite{SNT,sachdev,BMASR}:
\begin{equation}
C_\lambda(t)\sim |\Delta_\lambda|^{1/4}K_0(i|\Delta_\lambda|t) ,\quad \tilde
C_\lambda(t)\sim |\Delta_\lambda|^{1/4}  \label{C_tildeC}
\end{equation}
[with $K_0(z)$ the modified Bessel function]. In the quantum critical regime
($|\Delta_d|\ll T$), $C_d(t)\sim \tilde C_d(t)\sim t^{-1/4}$.

Employing Eqs. (\ref{chi_+}), (\ref{C_tildeC}), it is possible to evaluate
the retarded correlation function and thus $V(I)$ in either side of the
quantum critical point of the Ising model $d$. Below we show that the
resulting dramatically distinct behavior of the dissipative transport in the
disordered and ordered phases of the Ising system identifies them as
``superconducting" and ``insulating", respectively. \bigskip

\subsection{\textbf{Superconducting phases}}

\label{superconducting_phase} We first derive expressions for the $I-V$
characteristics near commensurate fields $B_{N}$ [Eq. (\ref{new_q})] where $
\Delta _{d}\sim\Delta _{d}^{(0)}>0$, in the low $T$ regime where Eq. (\ref{C_tildeC}) holds. Neglecting terms of order $e^{-\Delta
_{u}/T}$ and keeping the first order in $D$, we obtain form Eq. (\ref{33})
\begin{widetext}
\begin{equation}
%\begin{array}{lll}
V^{(1)}(I)=C\int^{\infty }_{-\infty } dt\sin \left(\frac{\pi It
}{2e}\right)\left( \frac{-\left(\pi \alpha T/v_{+}\right) }{\sinh (\pi T(t-i\varepsilon))
}\right)^{\frac{1}{K_{+}}} K_{0}(i\Delta_d t) \; ,\quad{\rm where}\quad
C\propto
DL|\Delta_u\Delta_d|^{1/4} .
%\end{array}
\label{49}
\end{equation}
\end{widetext}
For $\frac{\pi I}{2e}<\Delta_d$, this yields a non-linear $I-V$ curve
\begin{equation}
\begin{array}{l}
V^{(1)}(I) \approx V_s^{(1)}\sqrt{\frac{T}{\Delta_d(B)}}e^{-\Delta_d(B)/T
}\sinh \left( \frac{\pi I}{2eT}\right)\; , \\
V_s^{(1)}\propto D[\Delta_d(B)]^{K_+^{-1}(B)+\frac{1}{4}}
\end{array}
\label{IV1_SC}
\end{equation}
which exhibits a threshold at a critical current $I_c=\frac{2e\Delta_d}{\pi}$ in the limit
$T\rightarrow 0$.
In the Ohmic regime $I/e\ll T$, one obtains a contribution to the magnetoresistance of the form
\begin{equation}
\begin{array}{lll}
R^{(1)}(T,B)\approx R_s\sqrt{\frac{\Delta_d(B)}{T}} e^{-\Delta_d(B)/T}\; ,\\
R_s\propto D\left(\Delta_d(B)\right)^{K_+^{-1}(B)-\frac{3}{4}}\; .
\end{array}
\label{R_SC}
\end{equation}
Superimposed on a moderate monotonic increase with $B$ arising from $K_+(B)$ due
to the suppression of $\rho _{s}$ [Eq. (\ref{Kvg_c})], the exponential
factor leads to a strong \textit{decrease} and $R^{(1)}\rightarrow 0$ at $
T\rightarrow 0$ as long as $\Delta _{d}(B)>0$ is finite. The disordered
Ising phase is thus identified as \textit{superconducting}: it corresponds to a state where the phase of the SC order-parameter in the two wires is locked.
This suggests that the fields $\sigma _{d}$ physically represent phase-slips in the antisymmetric sector (which are gapped in this regime).

\begin{figure}
\includegraphics[width=1.0\linewidth]{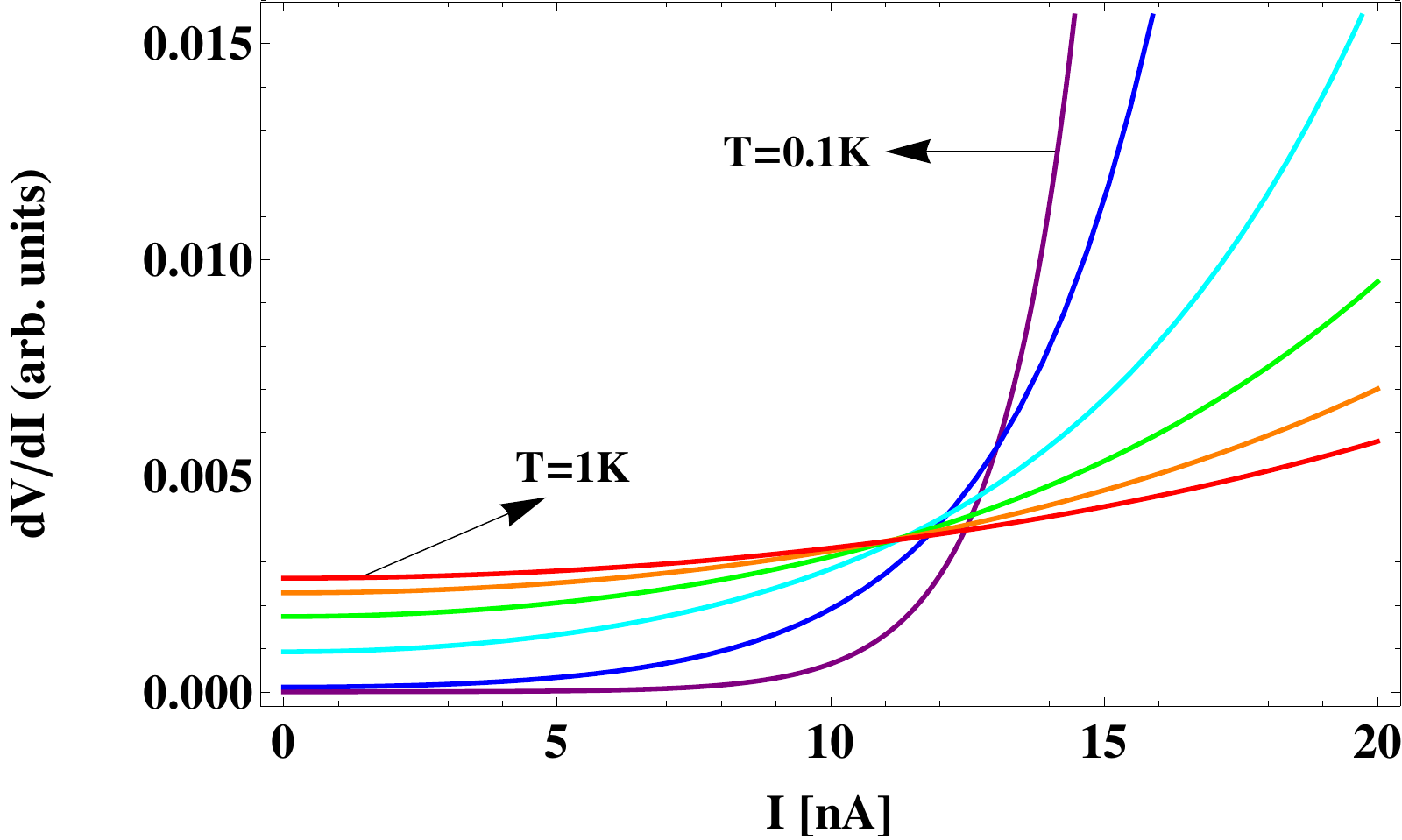}
\caption{(color online)
Differential resistance vs. current bias in the superconducting phase for temperatures $T=0.1$K, $0.2$K, $0.4$K, $0.6$K, $0.8$K and $1.0$K, for a fixed $B$ such that $\Delta_d=1.0$K and $K_+=2.1$ (see text); the disorder parameter is chosen such that $Da^3/v_-^2=0.1$.
  \label{fig:IVsc} }
\end{figure}

The above analysis indicates that the first order in $D$ yields an exponentially small voltage for $I,T\rightarrow 0$, suggesting that one should examine
the perturbation scheme in $H_D$ [Eq. (\ref{16})] more carefully \cite{OG2}. Indeed, if we evaluate the expectation value $\langle\,\rangle$ expanding to the next order in $D$, we find that the correlation functions $\chi_n$ acquire corrections to Eq. (\ref{34}) of the form
\begin{equation}
\delta\chi\propto D\langle
e^{\pm i2\theta_1(t)}e^{\pm i2\theta_2(t)}e^{\mp i2\theta_1(0)}e^{\mp i2\theta_2(0)}\rangle_0\; .
\label{deltaXi}
\end{equation}
Using Eq.(\ref{symmetric_antisymmetric}), this can be written as
\begin{equation}
\delta\chi\propto D\langle e^{i2\sqrt{2}\theta_+(t)}e^{-i2\sqrt{2}
\theta_+(0)}\rangle_0\; .  \label{deltaXi2}
\end{equation}
The resulting contribution to the voltage
\begin{equation}
V^{(2)}(I)\approx \frac{DL\pi}{4e}\overset{\infty}{\underset{-\infty}{\int}}
dt\sin \left(\frac{\pi It}{2e}\right)\delta\chi(t)  \label{54}
\end{equation}
is associated with scattering processes which do not involve the antisymmetric mode, and hence are not affected by the superconducting order. These correspond to coincidental events incorporating two scatterers located on two different wires simultaneously, and therefore their probability is of the order of $D^2$. The gapless symmetric mode experiences backscattering in such events, similarly to the usual plasmon mode in a strictly 1D SC wire.
Inserting the Luttinger liquid correlation function
\begin{equation}
\delta\chi(t)\sim D\left(\frac{-(\pi\alpha T/v_+)}{\sinh\{\pi
T\left(t-i\epsilon\right)\}}\right)^{\frac{4}{K_+}}\; ,  \label{55}
\end{equation}
we obtain \cite{GRbook}
\begin{widetext}
\begin{equation}
%\begin{array}{l}
V^{(2)}(I)=V^{(2)}_s\left\{B\left(-\frac{iI}{4eT}+ \frac{2}{K_+}, 1-\frac{4}{K_+}\right)- B\left(\frac{iI}{4eT}+ \frac{2}{K_+}, 1-\frac{4}{K_+}\right)\right\}
\quad{\rm where}\quad V_s^{(2)}\propto D^2
%\end{array}
\label{IV2_SC}
\end{equation}
\end{widetext}
and $B(x,y)=\frac{\Gamma(x)\Gamma(y)}{\Gamma(x+y)}$ is the Beta function.

The full $I-V$ characteristic in the SC phases ($\Delta_d(B)>T,I/e$) can finally be expressed as
\begin{equation}
V(I)= V^{(1)}(I)+V^{(2)}(I)\; ,
\label{IVSC_final}
\end{equation}
where $ V^{(1)}(I)$, $ V^{(2)}(I)$ represent contributions from odd and even orders in the disorder parameter $D$ respectively, and can be viewed as two resistors connected in series. To leading order in $D$, they are given by Eqs. (\ref{IV1_SC}) and (\ref{IV2_SC}), yielding the $I,T$-dependence depicted in Fig. 2. Note that although the second term is higher order in the scattering rate $D$, it becomes the dominant contribution in the limits $T,I\rightarrow 0$ as the first term is exponentially suppressed.
For $I/e\gg T$, this indicates a power-law $I-V$ relation
\begin{equation}
V(I)\sim  D^2I^{\kappa(B)+1}\; ,\quad \kappa (B)\equiv \frac{4}{K_+(B)}-2
\label{57}
\end{equation}
and in the Ohmic regime ($I/e\ll T$)
\begin{equation}
R(T,B)\sim D^2T^{\kappa(B)}\; .
\label{R2vsT}
\end{equation}
By definition of the Luttinger parameters $K_\pm$ [Eq. (\ref{Kvg_c})], $K_+\gtrsim K_-$ and hence the assumption $K_-=2$ implies $K_+\gtrsim 2$. As a consequence, the exponent $\kappa(B)$ [Eq. (\ref{57})] is small and slightly \textit{negative}. We therefore conclude that in spite of the phase--locking ordering of the antisymmetric phase mode, the true $T,I\rightarrow 0$ behavior of the electric transport exhibits an insulating behavior. In practice, however,
the insulating character may be manifested only at extremely low $T$. At moderately low $T$, the sub-leading term $ V^{(1)}(I)$ is expected to be appreciable, and indicate a threshold at a critical current $I_c$, directly related to an activation gap in the Ohmic resistance [Eq. (\ref{R_SC})]:
\begin{equation}
\log \,R\sim \Delta_d=\frac{\pi I_c}{2e}\; .
\label{logR_Ic}
\end{equation}
The oscillatory nature of $\Delta_d(B)$ as $B$ is tuned through commensurate and incommensurate values should be reflected in the $B$-dependence of $I_c$, which is maximized at commensurate values $B_N$ and vanishes in the vicinity of incommensurate regimes $B\sim B_{N+\frac{1}{2}}$.

\subsection{\textbf{Insulating phases}}
We next consider the insulating phase, realized in the vicinity of incommensurate fields $B\sim B_{N+\frac{1}{2}}$ such that
$\Delta_d<0$. In this case, both species of Ising models $u$ and $d$ are in the ordered phase, and for $T\ll |\Delta_d|$ the correlation function
characterizing the antisymmetric mode is given up to exponentially small corrections by a constant
\begin{equation}
\chi_-(t)\sim |\Delta_u\Delta_d|^{1/4}\; .
\end{equation}
As a result, $\chi(t)=\chi_+(t)\chi_-(t)$ [Eq. (\ref{34})] is dominated by the Luttinger liquid correlations
[Eq. (\ref{chi_+})] characterizing the symmetric mode.
Keeping the leading order in $D$ in Eq.(\ref{33}), we thus find an expression for the $I-V$ characteristics of the form
\begin{widetext}
\begin{equation}
\begin{array}{l}
V(I)=V_i\left\{B\left(-\frac{iI}{4eT}+ \frac{1}{2K_+}, 1-\frac{1}{K_+}\right)- B\left(\frac{iI}{4eT}+ \frac{1}{2K_+}, 1-\frac{1}{K_+}\right)\right\}
\; ,\quad{\rm where}\quad V_i\propto D|\Delta_u\Delta_d|^{1/4}\; .
\end{array}
\label{IV_characteristics}
\end{equation}
\end{widetext}
Typical plots of the resulting dynamic resistance $dV/dI$ vs. $I$ are depicted in Fig. 3, indicating a zero-bias peak at $I\rightarrow 0$, in sharp distinction from the SC phase (Fig. 1). For $I/e\gg T$, we obtain a diverging power-law
\begin{equation}
V(I)\sim DI^{1-\gamma(B)}\; ,\quad \gamma(B)\equiv 2-\frac{1}{K_+(B)}
\end{equation}
and in the Ohmic regime ($\frac{I}{e}\ll T$)
\begin{equation}
R(T,B)\sim DT^{-\gamma(B)}\; . \label{R_I}
\end{equation}
\begin{figure}
\includegraphics[width=1.0\linewidth]{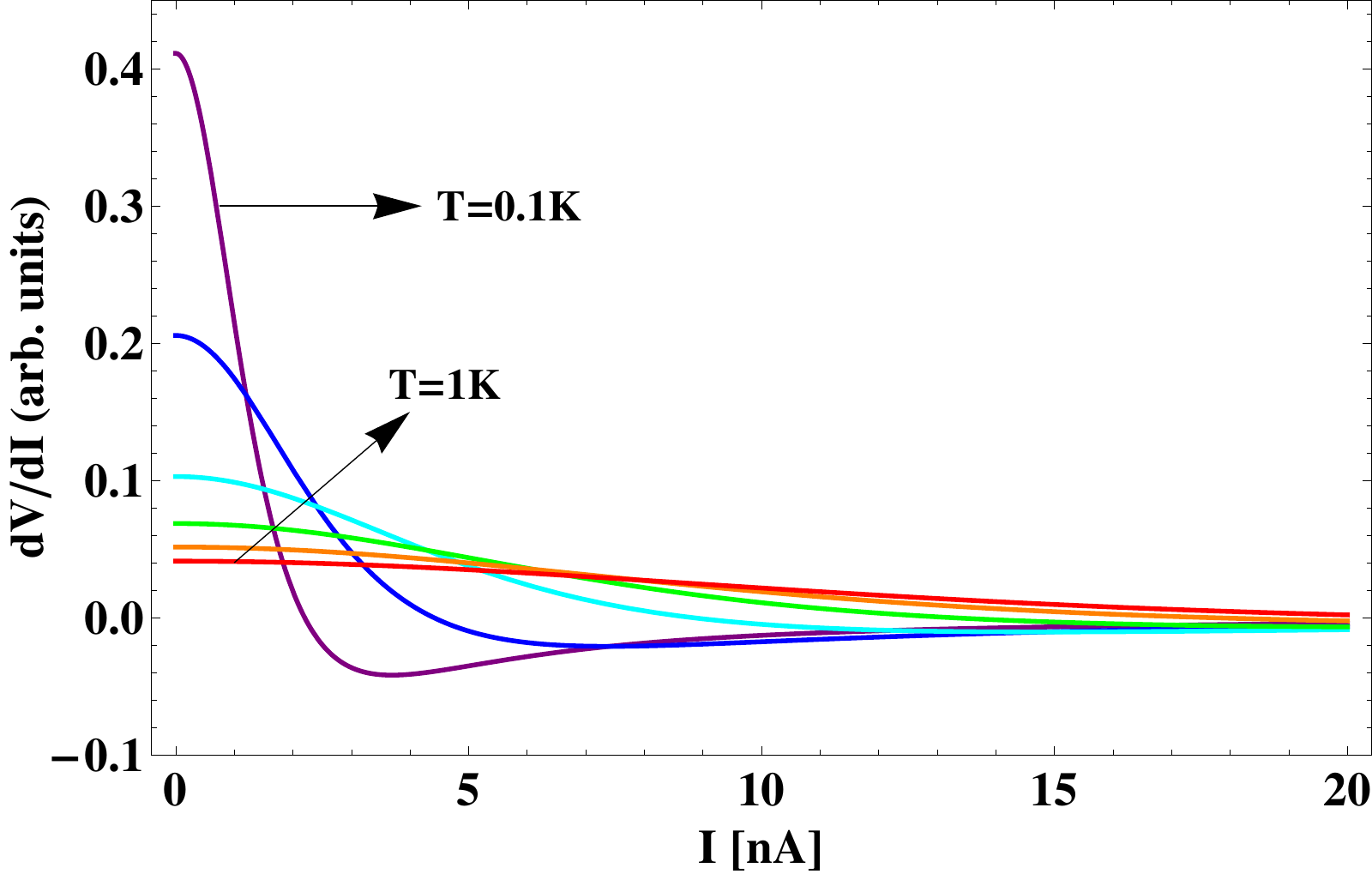}
\caption{(color online)
Differential resistance vs. current bias in the insulating phase for temperatures $T=0.1$K, $0.2$K, $0.4$K, $0.6$K, $0.8$K and $1.0$K, for a fixed $B$ such that $|\Delta_d|=1.0$K and $K_+=2.1$ (see text); the disorder parameter is chosen such that $Da^3/v_-^2=0.1$.
  \label{fig:IVins} }
\end{figure}
Compared to the power-law contributions to dissipation in the SC phase [Eqs. (\ref{57}) and (\ref{R2vsT})], these results indicate a stronger divergence at low $T$ and $I$. This behavior stems from the fact that the antisymmetric mode is in the \textit{insulating}, charge-ordered phase, and consequently backscattering processes by a single impurity are favored.
Moreover, since $K_+^{-1}\lesssim \frac{1}{2}$, the exponent $\gamma(B)>3/2$ indicating that the disorder potential is highly relevant.
In the truly $T,I\rightarrow 0$ limit (i.e., below a crossover temperature scale $T_{loc}$ which depends on the disorder strength $D$), the perturbative treatment of $H_D$ leading to Eq. (\ref{IV_characteristics}) is not valid and
localization takes over, yielding an exponentially diverging resistance \cite{book}. We note that at moderately low $T$ and $I$, Eq. (\ref{IV_characteristics}) is still valid and appears to be compatible with the experimental data \cite{shahar}.

\subsection{\textbf{Critical regime}}

The above analysis implies that the quantum critical points at $B_c^{(N)}$
(where $\Delta_d=0$) correspond to SC-I and I-SC transitions alternately, associated with the change of ordering in the antisymmetric mode from phase-ordered to charge-ordered ground state. These transitions are marked by a dramatic qualitative change in the shape of the non-linear $I-V$ curves, and in the $T$-dependence of the Ohmic resistance, as $B$ crosses $B_c^{(N)}$. However, note that unlike the 2D SIT, the quantum critical points can not be easily identified in the transport properties, e.g. as crossing points of isotherms where $R(B,T)$ exhibits a metallic behavior. In the critical regime ($T\gg |\Delta_d|$), the antisymmetric mode is characterized by power-law correlations $\chi_-(t)\sim t^{-1/4}$ and consequently
\begin{equation}
R(T,B)\sim T^{\frac{1}{4}-\gamma(B)}\; .
\label{R_c}
\end{equation}
This reflects once again an insulating behavior, characteristic to the 1D nature of the system. It stems from the presence of a gapless mode (the symmetric plasmon), which is not immune to backscattering processes.

\section{Discussion}
\label{sec:summary}

In this study, we have shown that the low-$T$ transport properties
of a ladder--like superconducting device subject to a
perpendicular magnetic field may signify a multitude of quantum
phase transitions from a SC to insulating phases alternately, when
its parameters are tuned close to the 2D SIT. These transitions
stem from the quantum mechanical nature of the vortex chain
accommodated along the central axis of the device, and reflect the
competition between a Josephson coupling and a charging energy
between the SC edges of the device, which govern the antisymmetric
phase--charge mode. The former dominates near commensurate values
of the vortex density, and the latter near incommensurate
($1/2$-integer) densities. The quantum critical points are of the
Ising type: this is a manifestation of the $Z_2$ symmetry
characterizing the antisymmetric mode, associated with
interchanging the two legs of the ladder.

The analysis presented in Sec. \ref{sec:transport} indicates,
however, that the electric transport properties are complicated by
the presence of a gapless symmetric phase--charge mode, which
provides a dissipative environment. As a result, the voltage
response to a current bias does not exhibit a strictly
superconducting behavior even in the phases classified as SC.
Nevertheless, for weakly disordered systems it is possible to
observe a clear signature of the SC nature of these phases at
finite $T$ and $I$. Subtracting the contribution of backscattering
exclusive to the symmetric mode, which can be viewed as a resistor
connected in series, one obtains an activated behavior of the
$I-V$ curve and the $T$-dependent resistance [see Fig. 1 and Eq.
(\ref{logR_Ic})]. This behavior is sharply distinct from the
insulating phases, where the differential resistance $dV/dI$
exhibit a zero-bias anomaly peak [see Fig. 2]. Moreover, in
principle it is possible to detect the quantum critical points
($B_c^{(N)}$) separating the two phases by probing the
$B$-dependence of the activated gap [Eq. (\ref{logR_Ic})].

It should be noted that the analysis thus far relies on some
crucial simplifying assumptions. In particular, it has been
assumed that the model for the antisymmetric mode is tuned to a
self-dual point, where $K_-=2$. In this special point, where both
the phase and charge fields are not well-defined, the chain of
vortices is exactly describable in terms of free Fermions. The
question arises, to what extent our results are robust against a
finite detuning away from the self-dual point, i.e. when
$K_-=2+\delta K$. Such corrections induce interactions among the
Fermions. However, since in both the SC and insulating phases the
Fermions are massive and excitations are gapped, these
interactions can be treated perturbatively as long as
$(v_-/a)\delta K\ll |\Delta_d|$. This approximation fails when $|\Delta_d|\rightarrow 0$ and the critical point is
shifted, but the Ising-type nature of the transition is maintained \cite{SAR}. The phenomenology manifested by the transport
properties as discussed above would therefore be essentially the same.

Another point of concern when adapting the model to describe a
realistic system is the role of finite size effects. In Sec.
\ref{sec:transport}, the correlation functions were evaluated for
finite $T$ assuming that the length of the system $L\rightarrow
\infty$. However, we note that the SC nanowires studied, e.g., in Ref.
\onlinecite{shahar}, typically have a finite length of the order of a
few microns. This introduces an additional low-energy cutoff
$T_L\equiv v_-/L$. Using typical values of the plasma velocity for $v_-$ (see, e.g., Ref. \onlinecite{zaikin}), we estimate $T_L\sim 1$K. This implies that for sub-Kelvin temperatures, $T_L$ effectively replaces $T$ as the low-energy cutoff. In the SC phases, the activated contribution to the resistance is therefore expected to be $\sim e^{-\Delta_d/T_L}$. Noting that $T_L$ is also associated with the zero-point energy of phase-fluctuations, this represents contribution due to macroscopic quantum tunneling of vortices out of a metastable state in the finite-size SC device \cite{LegCal}.

Finally, we wish to point out that a ladder--like SC device where the parameters are conveniently tunable (e.g., a Josephson ladder) can serve as an interesting playground for the study of emergent fractional degrees of freedom. In particular, when the gap $\Delta_d$ vanishes, the eigenstates of Eq. (\ref{H_-_majo}) (at zero energy) become Majorana Fermions. Therefore, as recently proposed by Tsvelik \cite{Majorana}, inhomogeneous SC devices can be potentially utilized to realize localized Majorana modes at interfaces between superconducting and insulating segments.

\acknowledgements

We thank T. Giamarchi, P. Goldbart, D. Pekker, Gil Refael, A.
Tsvelik and especially D. Shahar for useful discussions. E. S. is
grateful to the hospitality of the Aspen Center for Physics. This
work was supported by the US-Israel Binational Science Foundation
(BSF) grant 2008256 and the Israel Science Foundation (ISF) grant
599/10.


\begin{thebibliography}{99}
\bibitem{SITrev} For a review and extensive references, see A. F. Hebard, in
\textit{Strongly Correlated Electronic Materials} (The Los Alamos Symposium
1993), Eds. K. S. Bedell, Z. Wang, D. E. Meltzer, A. V. Balatsky and E.
Abrahams, Addison Wesley (1994), p. 251; G. T. Zimanyi, \textit{ibid} p.
285; Y. Liu and A. M. Goldman, Mod. Phys. Lett. B \textbf{8}, 277 (1994); S.
L. Sondhi, S. M. Girvin, J. P. Carini and D. Shahar, Rev. Mod. Phys. \textbf{
69}, 315 (1997) A. M. Goldman and N. Markovic, Physics Today \textbf{51}, 39
(1998).

\bibitem{SIT1D} K.Yu.Arutyunov, D. S. Golubev and A. D. Zaikin, Physics
Reports \textbf{464}, 1 (2008), and refs. therein.

\bibitem{LAMH} J. S. Langer and V. Ambegaokar, Phys. Rev. \textbf{164}, 498
(1967); D. E. McCumber and B. I. Halperin, Phys. Rev. B \textbf{1}, 1054
(1970).

\bibitem{TAPS} See, e.g., R. S. Newbower, M. R. Beasley and M. Tinkham,
Phys. Rev. B \textbf{5}, 864 (1972).

\bibitem{Giordano} N. Giordano, Phys. Rev. Lett. \textbf{61}, 2137 (1988);
N. Giordano, Phys. Rev. B \textbf{41}, 6350 (1990).

\bibitem{zaikin} A. D. Zaikin, D. S. Golubev, A. van Otterlo and G. T.
Zimanyi, Phys. Rev. Lett. \textbf{78}, 1552 (1997).

\bibitem{QPT2} S. Sachdev, \textit{Quantum Phase Transitions} (Cambridge
University Press (1999)).

\bibitem{Fisher} M. P. A. Fisher, Phys. Rev. Lett. \textbf{65}, 923 (1990).

\bibitem{shahar} A. Johansson, G. Sambandamurthy, N. Jacobson, D. Shahar,
and R. Tenne, Phys. Rev. Lett. \textbf{95}, 116805 (2005); A. Johansson, G.
Sambandamurthy and D. Shahar, unpublished.

\bibitem{glazman} C. Bruder, L.I. Glazman, A.I. Larkin, J.E. Mooij and A.
van Oudenaarden, Phys. Rev. B \textbf{59}, 1383 (1999).

\bibitem{networks} M. D. Stewart Jr., A. Yin, J. M. Xu and J. M. Valles Jr.,
Phys. Rev. B \textbf{77}, 140501(R) (2008); I. Sochnikov, A. Shaulov, Y.
Yeshurun, G. Logvenov and I. Bozovic, Nature Nanotechnology \textbf{5}, 516
(2010).

\bibitem{PGR}
D. Pekker, G. Refael and P. Goldbart, Phys. Rev. Lett. \textbf{107}, 017002 (2011).

\bibitem{OG1} E. Orignac and T. Giamarchi, Phys. Rev. B \textbf{64}, 144515
(2001).

\bibitem{AS} Y. Atzmon and E. Shimshoni, Phys. Rev. B \textbf{83}, 220518(R)
(2011).

\bibitem{likharev} K. K. Likharev, Sov. Phys. JETP \textbf{34}, 906 (1972).

\bibitem{book} T. Giamarchi, \textit{Quantum Physics in One Dimension},
(Oxford, New York, 2004).

\bibitem{SNT} D. G. Shelton, A. A. Nersesyan and A. M. Tsvelik, Phys. Rev. B
\textbf{53}, 8521 (1996).

\bibitem{Tsvelik} A. M. Tsvelik, Phys. Rev. B \textbf{83}, 104405 (2011).

\bibitem{Ising} A. O. Gogolin, A. A. Nersesyan and A. M. Tsvelik, \textit{
Bosonization and Strongly Correlated Systems} (Cambridge University Press,
1998).

\bibitem{sachdev} S. Sachdev and A. P. Young, Phys. Rev. Lett. \textbf{78},
2220 (1997).

\bibitem{BMASR} Decoupling of the $u$ and $d$ sectors is justified by the
significant difference in their masses ($\Delta _{u}\gg |\Delta _{d}|$); see
E. Boulat, P. Mehta, N. Andrei, E. Shimshoni and A. Rosch, Phys. Rev. B \textbf{76}, 214411 (2007).

\bibitem{OG2}
E. Orignac and T. Giamarchi, Phys. Rev. B \textbf{57}, 11713 (1998).

\bibitem{GRbook}
I. S. Gradshteyn and I. M. Ryzhik, \textit{Tables of Integrals, Series and Products} (Academic Press, 1980).

\bibitem{LegCal}
A. O. Caldeira and A. J. Leggett, Ann. Phys. \textbf{149}, 374 (1983), and references therein.

\bibitem{SAR}
E. Sela, A. Altland and A. Rosch, Phys. Rev. B \textbf{84}, 085114 (2011).

\bibitem{Majorana}
A. M. Tsvelik, arXiv:1106.2996.

\end{thebibliography}
\end{document}